\documentclass[12pt]{iopart}
\usepackage[english]{babel}
\usepackage{enumerate}
\usepackage{multicol}
\usepackage{fancyhdr}
\usepackage{amssymb}%,amsmath}
\usepackage{hhline}
\usepackage{wrapfig}
\usepackage{caption}
\usepackage{graphicx}% Include Fig. files
\usepackage{pifont}
\usepackage{makeidx}
\usepackage{cite}
\usepackage[T1]{fontenc}
\usepackage{epstopdf}
\usepackage{dcolumn}% Align table columns on decimal point
\usepackage{bm}% bold math
\usepackage{mathrsfs}
\usepackage{color}
\usepackage{booktabs}
\def\mxth{\mathsurround=0pt }
\def\xversim#1#2{\lower2.pt\vbox{\baselineskip0pt \lineskip-.5pt
  \ialign{$\mxth#1\hfil##\hfil$\crcr#2\crcr\sim\crcr}}}

\usepackage{epsf,epsfig}

\newcommand{\be}{\begin{equation}}
\newcommand{\ee}{\end{equation}}
\newcommand{\bea}{\begin{eqnarray}}
\newcommand{\eea}{\end{eqnarray}}

\newcommand{\ba}{\begin{array}}
\newcommand{\ea}{\end{array}}
\renewcommand{\bi}{\begin{itemize}}
\newcommand{\ei}{\end{itemize}}

\begin{document}
\nocite{*}
\title{Hyperfine and crystal field interactions in multiferroic HoCrO${_3}$}
\author{C~M~N~Kumar$^{1,2,3,\dagger}$, Y~Xiao$^{1,\ddagger}$, H~S~Nair$^{4}$, J~Voigt$^{1}$, B~Schmitz$^{1}$, T~Chatterji$^{5}$, N~H~Jalarvo$^{2,3}$, Th~Br\"{u}ckel$^{1}$}
\address{$^{1}$ J\"{u}lich Centre for Neutron Science JCNS and Peter Gr\"{u}nberg Institut PGI, JARA-FIT, Forschungszentrum J\"{u}lich, 52425 J\"{u}lich, Germany}
\address{$^{2}$ J\"{u}lich Centre for Neutron Science JCNS, Forschungszentrum J\"{u}lich GmbH, Outstation at SNS, Oak Ridge National Laboratory, Oak Ridge, Tennessee 37831, United States}
\address{$^{3}$ Chemical and Engineering Materials Division, Spallation Neutron Source, Oak Ridge National Laboratory, Oak Ridge, Tennessee 37831, United States}
\address{$^{4}$ Department of Physics, Colorado State University, Fort Collins, CO 80523, USA}
\address{$^{5}$ Institut Laue--Langevin, BP 156, F--38042 Grenoble Cedex 9, France}
\ead{$^{\dagger}$n.kumar@fz-juelich.de; naveenkumarcm@gmail.com, $^{\ddagger}$y.xiao@fz-juelich.de}

%\date{\today}

\begin{abstract}
We report a comprehensive specific heat and inelastic neutron scattering study to explore the possible origin of multiferroicity in HoCrO$_3$. We have performed specific heat measurements in the temperature range 100~mK--290~K and inelastic neutron scattering measurements were performed in the temperature range 1.5--200~K. From the specific heat data we determined hyperfine splitting at 22.5(2)~$\mu$eV and crystal field transitions at 1.379(5)~meV, 10.37(4)~meV, 15.49(9)~meV and 23.44(9)~meV, indicating the existence of strong hyperfine and crystal field interactions in HoCrO$_3$. Further, an effective hyperfine field is determined to be 600(3)~T. The quasielastic scattering observed in the inelastic scattering data and a large linear term $\gamma=6.3(8)$~mJmol$^{-1}$K$^{-2}$ in the specific heat is attributed to the presence of short range exchange interactions, which is understood to be contributing to the observed ferroelectricity. Further the nuclear and magnetic entropies were computed to be, $\sim$$17.2$~Jmol$^{-1}$K$^{-1}$ and $\sim$34~Jmol$^{-1}$K$^{-1}$, respectively. The entropy values are in excellent agreement with the limiting theoretical values. An anomaly is observed in peak position of the temperature dependent crystal field spectra around 60~K, at the same temperature an anomaly in the pyroelectric current is reported. From this we could elucidate a direct correlation between the crystal electric field excitations of Ho$^{3+}$ and ferroelectricity in  HoCrO$_3$. Our present study along with recent reports confirm that HoCrO$_3$, and $R$CrO$_3$ ($R=$ Rare earth) in general, possess more than one driving force for the ferroelectricity and multiferroicity.
\end{abstract}
\maketitle
%
%
%=============================================================================================%
\section{Introduction}
%=============================================================================================%
%
Perovskite chromites $R$CrO$_3$, where $R$ is a rare earth element or yttrium are revisited in the recent years as possible multiferroic materials in which multiple ferroic orders such as ferroelectricity and antiferromagnetism coexist as discussed below~\cite{CRSerrao_PRB2005,JRSahu_JMC2007,YSu_29JRareEarths2011,AGhosh_3JMCC2015AtypicalMultiferroicity,NABenedek_195JSSC2012PolarOctahedral,BRajeswaran_86PRB2012FieldInduced,KRSPMeher_89PRB2014PbservationOfElectric,ATApostolov_29MPLB2015MicroscopicApproach}. Compared to perovskite manganites, which are well studied in light of multiferroicity, the microscopic physical properties of chromites are not explored in detail and the mechanism for multiferroicity in HoCrO$_3$ is still under debate. The coexistence of ferroelectric and magnetic orders in rare-earth orthochromites was first suggested by Subba Rao $et~al.$~\cite{GVSubbaRao_SSC1968}. Based on dielectric studies it is reported that the heavy rare earth chromites, $R$CrO$_3$ ($R$=Ho, Er, Yb, Lu) undergo a ferroelectric transition in the temperature range $439-485$~K~\cite{JRSahu_JMC2007}. In a recent article, electrical polarization and magnetodielectric effect studies are reported for polycrystalline LuCrO$_3$ and ErCrO$_3$~\cite{KRSPMeher_89PRB2014PbservationOfElectric}. Although both LuCrO$_3$ and ErCrO$_3$ showed the presence of a polar state induced by magnetic ordering below $T_{\mathrm{N}}$, polarization was not affected by applied magnetic fields, so that the magnetoelectric coupling was not evident in these compounds. Further, the magnetodielectric effect observed in the case of ErCrO$_3$ is one order of magnitude higher compared with LuCrO$_3$ reflecting the role of different magnetism of rare-earth cations in ferroelectricity~\cite{KRSPMeher_89PRB2014PbservationOfElectric}.

Recently, Ghosh \textit{et al}., have studied the ferroelectric properties of polycrystalline HoCrO$_3$ by measuring the thermal variation of pyroelectric current~\cite{AGhosh_3JMCC2015AtypicalMultiferroicity}. It was found that pyroelectric current exhibits its maximum value around the antiferromagnetic transition temperature $T_{\rm N} = 140$~K, nevertheless, the ferroelectric order temperature which associated with the emergence of spontaneous electric polarization is observed at a higher temperature of $T\approx240$~K. The atypical multiferroic behavior observed in HoCrO$_3$ is argued to be a result of Ho displacements and oxygen octahedral rotations in the non--centrosymmetric $Pna2_1$ space group~\cite{AGhosh_3JMCC2015AtypicalMultiferroicity,NABenedek_195JSSC2012PolarOctahedral,AGhosh_107EPL2014PolarOctahedral,AIndra_28JPCM2016MagnetoelectricEffect}. The role of the rare earth ion in determining the physical properties of chromites $R$CrO$_3$ was revealed in a recent communication where the origin of ferroelectricity in orthochromites has been attributed to the instability of the symmetric position of the rare earth ion~\cite{BRajeswaran_86PRB2012FieldInduced}. The interaction between magnetic rare earth and week ferromagnetic Cr$^{3+}$ ions is the driving force for the breaking of symmetry, and thus the emergence of multiferroic behavior in these systems~\cite{BRajeswaran_86PRB2012FieldInduced,ATApostolov_29MPLB2015MicroscopicApproach}. Despite a debatable multiferriocity, orthochromites possess a plethora of physical phenomena, providing excellent opportunities to study and understand the basic interactions in materials. The detailed knowledge on the properties of rare earth ion is of particular importance to understand the multiferroicity in rare earth chromites $R$CrO$_3$.

The rare earth orthochromites crystallize in a distorted orthorhombic perovskite structure with four formula units per unit cell~\cite{SGeller_ActaCryst1956,SGeller_JChemPhys1956,EFBertaut_JPhysRad1956}. In HoCrO$_3$ the exchange coupling between the Cr$^{3+}$ nearest neighbors is predominantly antiferromagnetic and they order magnetically below the N\'{e}el temperature of $T_{\rm N}=140$~K~\cite{EFBertaut_2IEEE1966etude}. On the other hand, earlier reports differ on the aspect of Ho-ordering. Cooperatively induced ordering of Ho in HoCrO$_3$ was reported at 12~K~\cite{EFBertaut_2IEEE1966etude,PPataud_31JdePhys1970} whereas no ordering was observed by Hornreich $et. al.$, down to $1.5$~K~\cite{RMHornreich_2IntJMagn1972Magnetism}. Ferroelectricity is observed in $R$CrO$_3$ systems only when the $R$ ion is magnetic. This directly suggests the exchange interaction between Cr$^{3+}$ and $R$ is very important in inducing polarization and warrants the study of the local distortions around the $R$ ion as well as its magnetic properties. Hence, we have chosen HoCrO$_3$ as our subject to investigate the role of rare earth in the magnetic and thermodynamic properties of chromites.
%
%%
%%%
\section{Experimental details}
%%%
%%
%
Polycrystalline HoCrO$_3$ was synthesized by solid state reaction of Ho$_{2}$O$_{3}$ (3N) and Cr$_{2}$O$_{3}$ (4N) in stoichiometric  ratio. The precursors were mixed intimately and subsequently heat treated at 1100~${}^{\circ}$C for $48$~h. Then, the material was reground and annealed again at $1200~{}^{\circ}$C for $24$~h. The phase purity of the synthesized powder sample was confirmed by powder x-ray diffraction (PXRD) with Cu-$K{_{\alpha 1}}$ ($\lambda  = 1.54059~{\mbox{\AA}}$) radiation, using a Huber x-ray diffractometer (Huber G$670$) in transmission Guinier geometry. The profiles of the PXRD data were analyzed using the Rietveld method~\cite{HMRietveld_2JAC1969profile} implemented in the $FullProf$ software suit~\cite{JRCarvajal_192PB1993recent}. This confirmed the formation of orthorhombic single phase. The powder was then pressed into pellets and sintered at $1000~{}^{\circ}$C for $10$~h for further magnetic and thermal characterization. The heat capacity was measured in the temperature range from $100$~mK to $290$~K using a commercial Quantum Design Physical Property Measurement System equipped with a dilution insert. The heat capacity values were extracted using the relaxation method~\cite{JSHwang_68RevSciInstrum1997Measurement}. The background heat capacity of the microcalorimeter and the Apiezon $N$ grease used for thermal conduction was measured before the sample measurement and subtracted from the raw data to obtain the absolute heat capacity of the sample.

Inelastic neutron scattering (INS) measurements were carried out on the BASIS backscattering spectrometer of the Spallation Neutron Source (SNS), Oak Ridge National Laboratory, USA~\cite{EMamotov_82RevSciInstrm2011A-time-of-flight}. More details of this measurement is presented in reference~\cite{TChatterji_25JPCM2013Direct}. Inelastic neutron scattering experiments were also carried out on the high--resolution time--of--flight spectrometer FOCUS at the Spallation Neutron Source SINQ at PSI, Villigen in Switzerland. The polycrystalline sample was enclosed in an aluminum cylinder ($12$~mm diameter, $\sim$$45$~mm height) and placed into a He -- cryostat and the spectrum was collected using an incident energy of $19.61$~meV. Additional experiments were performed for the empty container as well as for vanadium to allow the correction of the raw data with respect to background, detector efficiency, absorption and detailed balance according to standard procedures. Inelastic data reduction and analysis was carried out using the software DAVE~\cite{RAzuah_114JResNatlInstStanTechnol2009Dave}.

%%
%%%
\section{Results and discussion}
%%%
%%
\subsection{Crystal Structure}
%
%%
%%%
\begin{figure}[!t]
\centering
\includegraphics[scale=0.52]{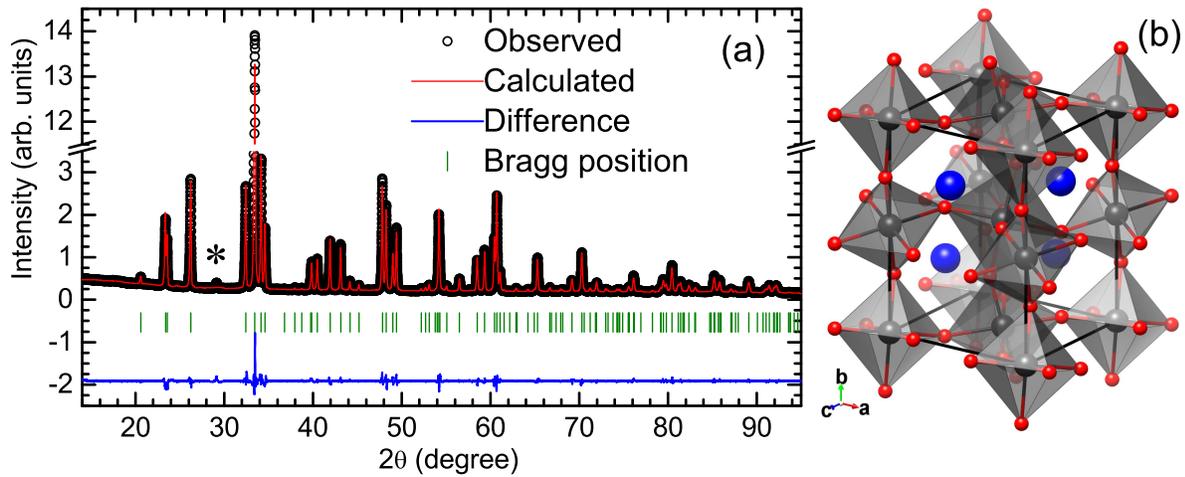}
\caption{(Color online) Observed (black circle) and calculated (red curve) PXRD patterns and their difference (blue curve) at $300$~K. The vertical bars denote the position of Bragg reflections. A spurious peak indicated by asterisk symbol could not be indexed. (b) Graphical representation of the crystal structure of HoCrO$_3$ with Cr-O$_6$ octahedra in $Pbnm$ space group. Blue, gray and red spheres indicate Ho, Cr and O atoms, respectively.}
\label{hco_XRD_RefStru}
\end{figure}
%%%
%%
%
%
The Rietveld refinement of room temperature PXRD data is presented in Fig.~\ref{hco_XRD_RefStru}(a), and the results are tabulated in table~1. Using Shannon--radii values for the ions~\cite{RDShannon_32ActCry1976revised} the value of the tolerance factor $t_{\rm G}$ for HoCrO$_3$ is found to be $\sim$$0.866$. Accordingly HoCrO$_3$ is an orthorhombically distorted perovskite with the space group $Pbnm$~\cite{SGeller_ActaCryst1956,SGeller_JChemPhys1956,EFBertaut_JPhysRad1956}. The crystal structure at $300$~K is in good agreement with the previously determined orthorhombic structure with similar lattice parameters~\cite{SQAmbrunaz_204BSFMCry1963, AGhosh_3JMCC2015AtypicalMultiferroicity}. The unit cell parameters obey the relationship, $a < c/\sqrt 2  < b$ which is characteristic of $O$--type orthorhombic structures. A buckling of the network of octahedra corresponding to cooperative rotation about a $\left[ {110} \right]$--axis leads to the $O$--type orthorhombic structure. The clinographic view of the CrO$_6$ octahedra in HoCrO$_3$ is presented in Fig.~\ref{hco_XRD_RefStru}(b). In perovskite manganites $R$MnO$_3$ ($R$ = La, Pr, Nd, Sm, Eu, Gd, Tb and Dy), in addition to the distortion due to buckling of the MnO$_6$ octahedron, a second distortion also arises because of the Jahn--Teller effect. This is because the Mn$^{3+}$ in $R$MnO$_3$ with four unpaired electrons the  $d$--shell in high spin state is Jahn--Teller active. On the other hand Cr$^{3+}$ in chromites with three unpaired electrons in $d$-orbitals is Jahn--Teller inactive. Thus a contribution to the lattice distortion in $R$CrO$_3$ due to the Jahn--Teller effect is ruled out.
\begin{table}[!htb]
    \centering
    \label{HCO_XRD_RoomTempStrParam}
    \caption[Room temperature structure parameters and atomic positions]{{Atomic positions, unit cell dimensions and discrepancy factors of HoCrO$_3$ obtained from the Rietveld refinement of the PXRD pattern at $300$~K. The values inside the brackets are the standard deviations.}}
    \begin{tabular*}{\columnwidth}{@{\extracolsep{\fill}}ccccc}
        \hline\hline\vspace*{-0.5cm}\\
        Atoms &        x        &       y       &       z      & B$_{iso}$~(\AA$^2$) \vspace*{-0.1cm}\\
         &					&				&				&   \\
        \hline\vspace*{-0.3cm}\\
        Cr   &         $0.5$           &    0          &      0        & 0.815(2) \\
        Ho   &         $-0.0168(8)$    &  $0.0655(5)$  &    $0.25$     & 1.031(3)\\
        O1   &         $ 0.1026(8)$    &  $0.4664(8)$  &    $0.25$     & 1.037(5)\\
        O2   &         $-0.3054(6)$    &  $0.3055(6)$  &    $0.0497(4)$& 1.152(1)\\
        \hline\vspace*{-0.5cm}\\
    \end{tabular*}
    \begin{tabular*}{\columnwidth}{@{\extracolsep{\fill}}cccc}
        \multicolumn{4}{c}{Unit Cell Dimensions}\vspace*{-0.0cm}\\
        $a = 5.2463(1)~{\mbox{\AA}}$ & $b= 5.5177(8)~{\mbox{\AA}}$ & $c = 7.5412(8)~{\mbox{\AA}}$ & $V=218.299(8)~{\mbox{\AA}^3}$ \\
        \hline\vspace*{-0.5cm}\\
    \end{tabular*}
    \begin{tabular*}{\columnwidth}{@{\extracolsep{\fill}}cccc}
        \multicolumn{4}{c}{Discrepancy Factors}\vspace*{0.0cm}\\
        \vspace{0.2cm}
        $R_p=3.98\%$ &   $R_{wp}=5.53\%$ & $R_{exp}=3.06\%$ & $\chi^2=3.27$ \vspace*{-0.2cm}\\
        \hline
    \end{tabular*}
\end{table}
\subsection{Specific heat}
\begin{figure*}[!b]
\centering
\includegraphics[scale=0.3]{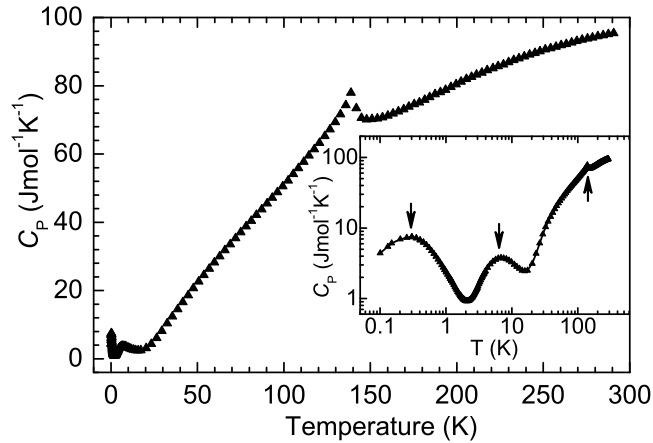}
\caption{(Color online) The temperature dependence of specific
    heat $C_{\rm{P}}$ of HoCrO$_3$. Three distinct features in the $C_{\rm{P}}$ are clearly visible in the inset.}
\label{HCO_SpecHeat_Full}
\end{figure*}

The variation of the specific heat ($C_{\rm P}$) of HoCrO$_3$ with temperature is presented in Fig.~\ref{HCO_SpecHeat_Full}. The low temperature features in $C_{\rm P}$ can be visualized clearly in a log-log plot as shown in the inset of Fig.~\ref{HCO_SpecHeat_Full}. The direct inspection of the specific heat curve evidences the presence of three main contributions or features: (1) a sharp increase in specific heat below $2$~K with a maximum at $\sim$$0.3$~K due to anomalously large hyperfine interaction between the electronic and nuclear spins of $^{165}$Ho$^{3+}$ leading to a nuclear-Schottky specific heat ($C_{\rm{N}}$); (2) the electronic Schottky contribution  ($C_{e}$) from thermal depopulation of the ${}^5I_8$ ground state multiplet of Ho$^{3+}$ with a maximum at $\sim$$7$~K; and (3) the $\lambda$--like anomaly with a peak at $T$$\approx$142~K due to the magnetic ordering of the Cr$^{3+}$ moments. To determine different contributions to $C_{\rm{P}}$, a detailed analysis was performed in two steps, first $C_{\rm{P}}$ was modeled in the temperature range $0.1~\mathrm{K}\leqslant{T}\leqslant{30}$~K and then $2~\mathrm{K}\leqslant{T}\leqslant{290}$~K.
%
%%
%%%
%%%
%%
%
\subsubsection{Specific heat in the temperature range $0.1~\textrm{K}\leqslant{T}\leqslant{30}~\textrm{K}$\\}
The specific heat below $30$~K has two broad features with maxima around $\sim$$0.3$~K and $\sim$$7$~K. From the low temperature specific heat measurements of Ho metal it was found that anomalously large hyperfine interaction between the electronic and nuclear spins of Ho commonly leads to a nuclear Schottky anomaly with a maximum at $\approx$$0.3$~K~\cite{HVKempen_30Physica1964}. The Hamiltonian for hyperfine interactions can be written in the form~\cite{BBleney_78PPS1961}:
\begin{equation}
\label{HCO_LT_HC_hamiltonian}
\frac{\mathscr{H}}{{{k_{\rm{B}}}}} = a'{I_z} + P\left[ {I_z^2 - {\frac{1}{3}} I\left( {I + 1} \right)} \right]
\end{equation}
where, $a'$ is the magnetic hyperfine constant, which is a measure of the strength of the hyperfine interaction between the nuclear moment and the magnetic moment associated with the 4$f$ electrons. $P$ is the quadrupolar coupling constant. The field is applied in the $z$ direction. Since the projection $I_z$ can take $2I+1$ values, i.e. $- I, - I + 1,...I$, the hyperfine specific heat $C_{\rm{P}}$  will be a Schottky type specific heat, associated with the $2I+1$ hyperfine levels. The nuclear spin of $^{165}$Ho with 100\% natural abundance is $I=7/2$, while $^{53}$Cr with $I=3/2$ has a natural abundance of 9.5\%. One can calculate the mean square of effective nuclear moment for the natural abundance of active isotopes of Ho and Cr, $\overline{\mu^2_{eff}}$, yielding $22.4~\mu^2_{\rm{N}}$ for $^{165}$Ho and $0.035~\mu^2_{\rm{N}}$ for Cr. With these effective nuclear moments one can conclude that the Cr hyperfine contribution is about three orders of magnitude smaller than the Ho one. Thus only the Ho contribution is taken into account in the
calculation of nuclear hyperfine contribution to the specific heat. Due to the hyperfine interaction the holmium nucleus has $2I+1=8$ possible spin orientations relative to an effective field {$H_{eff}$}. The energies ${{\varepsilon _i}}/k_{\rm{B}}$ of various nuclear spin states, i.e. the eigenvalues of the Hamiltonian in equation~(\ref{HCO_LT_HC_hamiltonian}) are
\begin{equation}
\label{HCO_Nuclear_Energy}
\frac{{{\varepsilon _i}}}{{{k_{\rm{B}}}}} = a'i + P\left[ {{i^2} - {\frac{1}{3}} I\left( {I + 1} \right)} \right]
\end{equation}
where, $i=-7/2,~-5/2,....,~5/2,~7/2$. Information about $a'$ and $P$ can be obtained by measuring the heat capacity at sufficiently low temperatures. In case of holmium, the quadrupolar coupling contribution is small and can be neglected~\cite{JRMignod_63PSSb1974}. Therefore for $P\approx0$ the equation~(\ref{HCO_Nuclear_Energy}) reduces to,
\begin{equation}
\label{HCO_Nuclear_Energy_approximate}
\frac{{{\varepsilon _i}}}{{{k_{\rm{B}}}}} \approx a'i
\end{equation}
%%%
%%
%
The specific heat  in the temperature range $0.1-30$~K is modeled by taking into consideration the contributions from nuclear specific heat $C_{\rm{N}}$, an electronic Schottky term $C_{e}$ and a lattice term $C_{\rm{L}}$. Thus at low temperatures the $C_{\rm{P}}$ of HoCrO$_3$ is given by,
\begin{equation}
\label{HCO_LTSpecHeat}
C_{\rm{P}}  = C_{\rm{N}} + C_{e} + C_{\rm{L}}
\end{equation}
\noindent
where, $C_{\rm{N}}$ and $C_{e}$ are given by the general expression for an n-level Schottky specific heat term given by~\cite{ATari_SpecificHeatBook2003}:
\begin{equation}
\label{HCO_Gen_Sch_Term}
{C_{Schottky}} = \frac{R}{{{T^2}}}\frac{{\sum\limits_i {\sum\limits_j {\left( {\Delta _i^2 - {\Delta _i}{\Delta _j}} \right)} } \exp \left[ { - \left( {{\Delta _i} + {\Delta _j}} \right)/T} \right]}}{{\sum\limits_i {\sum\limits_j {\exp \left[ { - \left( {{\Delta _i} + {\Delta _j}} \right)/T} \right]} } }}
\end{equation}
\noindent
In this expression, ${\Delta _i} = {{\varepsilon _i}}/k_{\rm{B}}$ and $R=8.314$~Jmol$^{-1}$K$^{-1}$ is universal gas constant. To calculate the nuclear specific heat $C_{\rm{N}}$, a Schottky curve for $^{165}$Ho with $I$=7/2, eight equally spaced energy levels with splitting energy, ${{{\varepsilon _i}} \mathord{\left/{\vphantom {{{\varepsilon _i}} {{k_{\rm{B}}}}}} \right.
 \kern-\nulldelimiterspace} {{k_{\rm{B}}}}} \approx a'i$, where $i =  - {\raise0.5ex\hbox{$\scriptstyle 7$}\kern-0.1em/\kern-0.15em\lower0.25ex\hbox{$\scriptstyle 2$}}, -{\raise0.5ex\hbox{$\scriptstyle 5$}\kern-0.1em/\kern-0.15em\lower0.25ex\hbox{$\scriptstyle 2$}},...{\raise0.5ex\hbox{$\scriptstyle 5$}\kern-0.1em/\kern-0.15em\lower0.25ex\hbox{$\scriptstyle 2$}}, {\raise0.5ex\hbox{$\scriptstyle 7$}\kern-0.1em/\kern-0.15em\lower0.25ex\hbox{$\scriptstyle 2$}}$, is used. To calculate the electronic Schottky specific heat $C_{e}$, a simple two level Schottky term with energy splitting ${{{\varepsilon _s}} \mathord{\left/{\vphantom {{{\varepsilon _s}} {{k_{\rm{B}}}}}} \right.
 \kern-\nulldelimiterspace} {{k_{\rm{B}}}}}$ is used, as the contribution from higher energy terms is negligible in the temperature range $0.1-30$~K. In this temperature range the lattice contribution is expressed by a single Debye term. At low temperatures, when $T<<\Theta _{\rm{D}}$, the Debye temperature, the Debye specific heat can be represented by well-known Debye~T$^3$--law as~\cite{ATari_SpecificHeatBook2003},
\begin{equation}
\label{HCO_LT_Debye}
{C_{Debye}} = R\frac{{234{r_{\rm{D}}}{T^3}}}{{\Theta _{\rm{D}}^3}} = \beta_3 {T^3}
\end{equation}
\noindent
where, $r_{\rm{D}}$ is the number of atoms per molecule.\\
%
%%
%%%
\begin{figure*}[!t]
    \centering
    \includegraphics[scale=0.295]{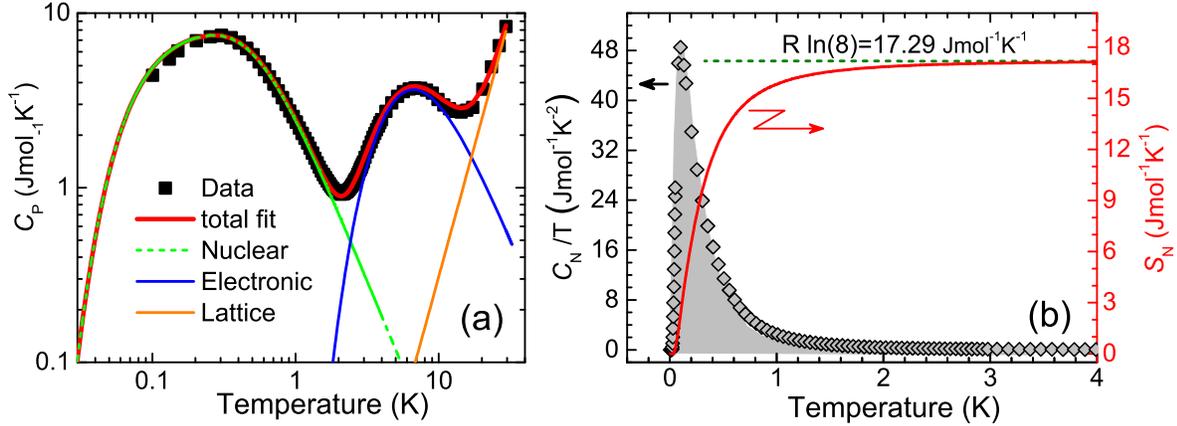}
    \caption{(Color online) (a) Double logarithmic plot of specific heat $C_{\rm{P}}$ measured at zero magnetic field plotted along with the refined model using equation~\ref{HCO_LTSpecHeat}. Different contributions to the total specific heat $C_{\rm{P}}$ are also shown. (b) A plot of $C_{\rm{N}}/T$ vs. T and the entropy associated with the nuclear specific heat $S_{\rm{N}}$ obtained by the numerical integration of $C_{\rm{N}}/T$ (shaded region) using equation~\ref{HCO_NucEntropy}. The horizontal dashed-line corresponds to the theoretical limiting value of the entropy.}
    \label{HCO_LT_SpHeat_Nuc_Entropy}
\end{figure*}
%%%
%%

The fit results to equation~(\ref{HCO_LTSpecHeat}) are given in Fig.~\ref{HCO_LT_SpHeat_Nuc_Entropy}~(a). From the fit the values of electronic Schottky splitting energy $\varepsilon_s$ and the Debye temperature ${\Theta _{\rm{D}}}$ are found to be, 1.379(5)~meV and $318(2)$~K, respectively. The Schottky energy 1.379(5)~meV is consistent with earlier reports based on specific heat measurements~\cite{PPataud_31JdePhys1970} and optical absorption Zeeman spectroscopy~\cite{RCourths_24ZPB1976optical}. The value of $a'$ from our analysis is found to be 0.2615(6)~K which is slightly smaller compared to the value found in metallic Ho, $a'\approx0.31-0.32$~K~\cite{JEGordon_124PR1961,OVLounasmaa_128PR1962,HVKempen_30Physica1964}, in inter-metallic HoCo$_2$, $a'\approx0.31$~K~\cite{DBloch_12SSC1973} and in paramagnetic salts, $a'\approx0.31$~K~\cite{JEGordon_124PR1961}. The energy difference between two adjacent nuclear levels due to hyperfine field thus calculated from equation~(\ref{HCO_Nuclear_Energy_approximate}) is $22.5(2)$~$\mu$eV. From inelastic neutron scattering measurements it is possible to observe these energy levels directly, as an inelastic peak centered at $\sim$$22.5$~$\mu$eV. Similar observations were made in spin--ice compound Ho$_2$Ti$_2$O$_7$ which shows a nuclear Schottky peak around $0.3$~K in the specific heat data~\cite{YMJana_61PRB2000}, which was later observed in inelastic neutron scattering measurements as a peak at $\sim$$26$~$\mu$eV~\cite{GEhlers_102PRB_2009}. The effective magnetic (hyperfine) field at the holmium nuclei can be computed by writing~\cite{OVLounasmaa_128PR1962},
\begin{equation}
\label{HCO_HypConst}
a' = \frac{{\mu {H_{eff}}}}{{{k_{\rm{B}}}I}}
\end{equation}
\noindent
where, $\mu=4.17~\mu_{\rm{N}}$ for $^{165}$Ho and $\mu_{\rm{N}}$ is the nuclear magneton ($\mu_{\rm N}=5.051\times10^{-27}$~JT$^{-1}$). Using the value of $a'$ in equation~(\ref{HCO_HypConst}) the hyperfine field,  $H_{eff}$ is found to be $600(3)$~T. This value is comparable to the values reported for Ho$_2$Ti$_2$O$_7$ (720~T)~\cite{YMJana_61PRB2000}, metallic Ho (770~T)~\cite{OVLounasmaa_128PR1962} based on specific heat data and in ErCrO$_3$ (530~T)~\cite{MEibschutz_178PR1969} based on M\"{o}ssbauer spectroscopy.

The nuclear specific heat $C_{\rm{N}}$ was obtained by subtracting the lattice and electronic contributions from $C_{\rm{P}}$. The entropy $S_{\rm{N}}$ associated with the nuclear specific heat was calculated by the numerical integration of $C_{\rm{N}}/T$. $S_{\rm{N}}$ is given by the expression,
\begin{equation}
\label{HCO_NucEntropy}
S_{\rm{N}}\left( T \right) = \int\limits_0^T {\left( {\frac{{{C_{\rm{N}}}}}{{T}}} \right)} dT
\end{equation}
\noindent
The $C_{\rm{N}}/T$ versus $T$ and the computed $S_{\rm{N}}$ are presented in Fig.~\ref{HCO_LT_SpHeat_Nuc_Entropy}(b). The nuclear entropy $S_{\rm{N}}$ reaches a maximum value of $\sim$$17.2$~Jmol$^{-1}$K$^{-1}$ at $\sim$$5$~K. The theoretical limiting value of entropy for $^{165}$Ho with nuclear spin $I=7/2$ is calculated as $Rln(2I+1)=Rln(8)\simeq17.29$~Jmol$^{-1}$K$^{-1}$. It is shown as an horizontal dashed line in Fig.~\ref{HCO_LT_SpHeat_Nuc_Entropy}(b). An excellent agreement between the experimental $S_{\rm{N}}$ with the theoretical value suggests that only contribution to the low temperature peak in the specific heat is from the nuclear Schottky term due to hyperfine interactions.

\subsubsection{Specific heat in the temperature range, $2~\textrm{K}\leqslant{T}\leqslant{290}~\textrm{K}$.\\}
\label{sec:HCO_HT_SpecHeat}
\begin{figure*}[!t]
\centering
\includegraphics[scale=0.57]{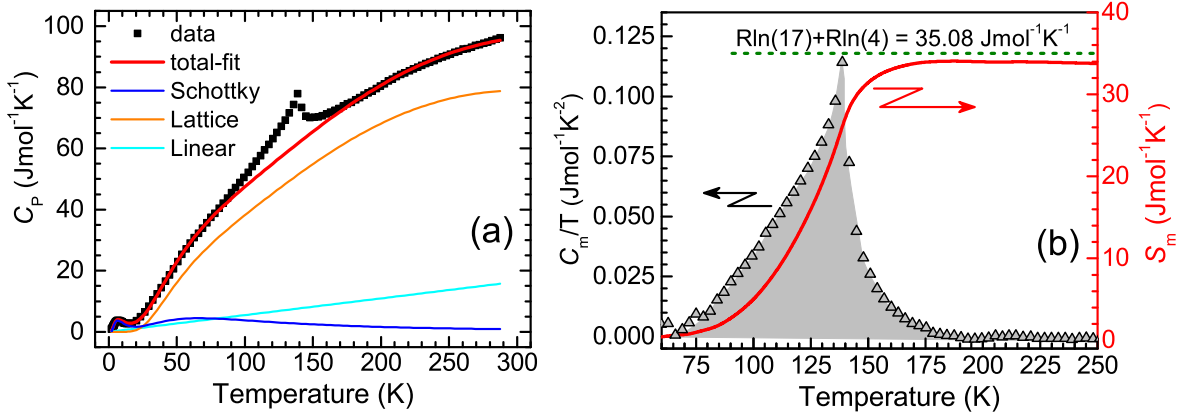}
\caption{(Color online) (a) Specific heat, $C_{\rm{P}}$ measured at zero magnetic field plotted along with the fitted results using the equation~(9). Different contributions to the total specific heat $C_{\rm{P}}$ are also shown. (b) The $C_{\rm{m}}$/T vs. T and the magnetic entropy $S_{\rm{m}}$, obtained by the numerical integration of $C_{\rm{m}}/T$ (shaded region). The horizontal dashed-line corresponds to the theoretical limiting value of the entropy.}
\label{HCO_Cp_HighT_Fit_HCO_Cm_Entropy}
\end{figure*}

The non-magnetic contribution to the specific heat $C_{nm}$ in the temperature region $2-290$~K is fitted assuming the contributions from an electronic Schottky term $C_{Schottky}$, lattice term $C_{Lattice}$ and a linear term $C_{Linear}$. HoCrO$_3$ has 5 atoms per formula unit, which implies that 15 vibrational modes to the phononic specific heat exist. Taking in to account this constraint, we approximate the lattice contribution to the specific heat ($C_{Lattice}$) as sum of a Debye term ($C_{Debye}$) and two distinct Einstein terms $C_{Einstein}$. The specific heat associated with the magnetic ordering, ($C_{\rm{m}}$) results in a $\lambda$--like transition with a maximum at $\sim$$142$~K. $C_{\rm{m}}$ is obtained by subtracting $C_{nm}$ from the experimental data ($C_{\rm{P}}$). The magnetic entropy $S_{\rm{m}}$ associated with $C_{\rm{m}}$ is obtained by the numerical integration of $C_{\rm{m}}/T$. The $C_{nm}$ and $C_{\rm{m}}$ in the temperature range $2-290$~K can be written as~\cite{ATari_SpecificHeatBook2003,ESRGopal_SpecificHeatBook1966},
\begin{eqnarray}
 && C_{nm} = C_{Schottky} + C_{Lattice} + C_{Linear} \label{Non-Magnetic}\\
 && C_{Lattice} =  C_{Debye}  +  C_{Einstein} \label{HT_SpecHeat}\\
 && C_{\rm{m}}=C_{\rm{P}} - C_{nm}
\end{eqnarray}
\noindent
Here $C_{Schottky}$ is purely electronic Schottky term as we are fitting only above 2~K, where the nuclear contribution is negligible. The non-magnetic contribution to the specific heat in this temperature is obtained by equations~(\ref{HCO_Gen_Sch_Term}), (\ref{Non-Magnetic}), (\ref{HT_SpecHeat}) and,
\begin{eqnarray}
&& C_{Debye}  =  9rR/x^{3}_{\rm{D}} \int^{x_{\rm{D}}}_{0}{x^4 e^{x}/(e^{x} - 1)^2 dx}\label{HCO_Debye}\\
&& C_{Einstein} = 3rR\sum\limits_i {a_i \left[ {{{x_i^2 e^{x_i } } \mathord{\left/ {\vphantom {{x_i^2 e^{x_i } } {\left( {e^{x_i }  - 1} \right)^2 }}} \right. \kern-\nulldelimiterspace} {\left( {e^{x_i }  - 1} \right)^2 }}} \right]}\label{HCO_Einstein}\\
&& {C_{Linear}} = \gamma T\label{HCO_Linear}
\end{eqnarray}
\noindent
In these expressions, $R$ is the gas constant, $x_{\rm{D}}$ = $\hbar\omega_{\rm{D}}/k_{\rm{B}}T$, $x_i$ = $\hbar\omega_{\rm{E}}/k_{\rm{B}}T$, $k_{\rm{B}}$ is Boltzmann constant, $\gamma$ is the coefficient of the linear term and $r$ is the number of atoms per molecule. The fitting was performed excluding the data in the temperature range 60--200~K. The different contributions to the measured specific heat obtained from fitting are presented in Fig.~\ref{HCO_Cp_HighT_Fit_HCO_Cm_Entropy}(a). The values of Debye temperature $\Theta_{\rm D}$ and two Einstein temperatures ($\Theta_{E1}$, $\Theta_{E2}$) obtained from the fit are $538(14)$~K (46(1)~meV) and $784(15)$~K (67(1)~meV) and $176(2)$~K (15.17(2)~meV), respectively. This contribution to $C_{Lattice}$ was parametrized by using 3 Debye modes with Debye temperature ($\Theta_{\rm D}$), 7 Einstein modes with Einstein temperature $\Theta_{E1}$, and another 5 Einstein modes with $\Theta_{E2}$. Although the used parametrization over simplify the phonon spectrum, the obtained key results are not influenced by subtleties in the choice of the modeled lattice contribution, i.e., by the number of Debye and Einstein contribution or by the used absolute values within reasonable error bars~\cite{MSchapers_88PRB2013ThremodynamicProperties}.

Based on optical absorption spectroscopy~\cite{RCourths_24ZPB1976optical}, magnetization and magnetic susceptibility~\cite{RMHornreich_2IntJMagn1972Magnetism}, specific heat~\cite{PPataud_31JdePhys1970charleurs}, elastic neutron diffraction~\cite{NShamir_90PhysicaBC1977magnetic} and inelastic neutron scattering (INS)~\cite{NShamir_90PhysB1977inelastic} experiments five electronic Schottky levels were observed in HoCrO$_3$ with the fifth level being at $\sim$$272$~K~(23.4~meV). To calculate $C_{Schottky}$, a Schottky curve for five energy levels is used which contribute to $C_{Schottky}$ below $300$~K. The ground state energy level is assumed to be zero i.e., ${{\varepsilon _1}}/k_{\rm{B}}=0$. From the low temperature specific heat analysis, where a simple two level Schottky model was used to calculate $C_{Schottky}$, the energy splitting is found to be 1.379(5)~meV. We fixed this value as $\varepsilon_2$ in the present calculations. The higher energy levels are determined from the fit as $\varepsilon_{3}$=$10.37(4)$~meV, $\varepsilon_{4}$=$15.49(9)$~meV and $\varepsilon_5$=$23.44(9)$~meV. These values are in excellent agreement with the reported crystal field energy values in HoCrO$_3$ (within 0.5$\%$).

After subtracting the contributions, $C_{Schottky}$, $C_{Lattice}$ and $C_{Linear}$ from the total specific heat $C_{\rm{P}}$, the magnetic specific heat $C_{\rm{m}}$ is obtained, which can be seen as the deviation from the total fit in Fig.~\ref{HCO_Cp_HighT_Fit_HCO_Cm_Entropy}(a) in the temperature range $60-180$~K. The magnetic entropy, $S_{\rm{m}}$ was calculated by the numerical integration of $C_{\rm{m}}/T$ analogous to the expression (\ref{HCO_NucEntropy}), by replacing $C_{\rm{N}}$ by $C_{\rm{m}}$. The $C_{\rm{m}}/T$ versus $T$ and computed $S_{\rm{m}}$ is presented in Fig.~\ref{HCO_Cp_HighT_Fit_HCO_Cm_Entropy}(b). The experimental magnetic entropy value reaches a maximum value of $\sim$34~Jmol$^{-1}$K$^{-1}$ around 180~K, above $T_{\mathrm{N}}$. The theoretical limiting value of $S_{\rm{m}}$ was calculated by adding the contributions due to ordering of both Ho$^{3+}$ ($J=8$) and Cr$^{3+}$ ($S =3/2$) moments, i.e., $R\ln (17) + R\ln (4) \simeq 35.08$~Jmol$^{-1}$K$^{-1}$, indicated as a horizontal dotted line in Fig.~\ref{HCO_Cp_HighT_Fit_HCO_Cm_Entropy}(b). The experimental magnetic entropy is very close to the theoretical magnetic entropy around 180~K $(\sim$$97\%)$. The good agreement between the experimentally found value for the magnetic entropy and that calculated for the spin only component of Cr$^{3+}$ ions and orbital magnetic moment of Ho$^{3+}$ ions, allows for concluding the following: first, the orbital moments of Cr$^{3+}$ ions appear to be quenched while keeping the full spin moment. Second, the lattice contribution seems to be described sufficiently well by the Debye and Einstein models.

The value of linear coefficient $\gamma$ from the fit is found to be $6.3(8)$~mJmol$^{-1}$K$^{-2}$. This value of $\gamma$ is comparable to the reported values, $5-7$ ~mJmol$^{-1}$K$^{-2}$ which is associated with conduction electrons in some doped manganites~\cite{LGhivelder_189JMMM1998specific,JJHamilton_54PRB1996low,JMDCoey_75PRL1995Electron,BFWoodfield_78PRL1997Low}. The linear coefficient is usually attributed to charge carriers, and is proportional to the density of states at the Fermi level. However, HoCrO$_3$ is an electrical insulator, thus the origin of linear term should be interpreted with caution. Several electrical insulators have been reported with large values of $\gamma$, viz., LaMnO$_{3+\delta}$ ($\gamma \approx 20$~mJmol$^{-1}$K$^{-2}$) ~\cite{LGhivelder_60PRB1999}, BaVS$_3$ ($\gamma=15.7$~mJmol$^{-1}$K$^{-2}$) ~\cite{HImai_65JPSJ1996calorimetric} and La$_{2.3}$YCa$_{0.7}$Mn$_2$O$_7$ ($\gamma=41.5$~mJmol$^{-1}$K$^{-2}$)~\cite{PRaychaudhuri_10JPCM1998metal}. The origin of a linear contribution in these electrical insulators was attributed to a variety of magnetic phenomena. The most plausible explanation for the appearance of a linear term in the heat capacity in HoCrO$_3$ is due to disordered Ho$^{3+}$ as in the case of insulating Ho$_{1-x}$Y$_x$MnO$_3$. For this compound it was observed that with increasing Y content $\gamma$ is reduced and drops to zero at $x=0.9$, indicating the dependence of $\gamma$ on the Ho content~\cite{HDZhou_75PRB2007}. It was interpreted that the appearance of the linear term is due to high degeneracy of disordered Ho$^{3+}$ spins above the ordering temperature. From high resolution neutron spectroscopy a huge quasielastic scattering was observed in HoCrO$_3$, which was understood as due to fluctuating disordered Ho electronic moments~\cite{TChatterji_25JPCM2013Direct}. This supports our interpretation of disordered Ho electronic moments as a possible source of a linear term in the observed specific heat. It is worth noting at this point that, to model the low temperature specific heat data using equation~(\ref{HCO_LTSpecHeat}), a linear term was not required which can be understood due to spin ordering of Ho$^{3+}$ at these low temperatures~\cite{CMNKumar_Unpublished}. These observations confirm that the main origin of the linear term in HoCrO$_3$ is disordered Ho$^{3+}$ spins.
\subsection{Inelastic neutron scattering}
\begin{figure*}[!t]
\includegraphics[scale=0.54]{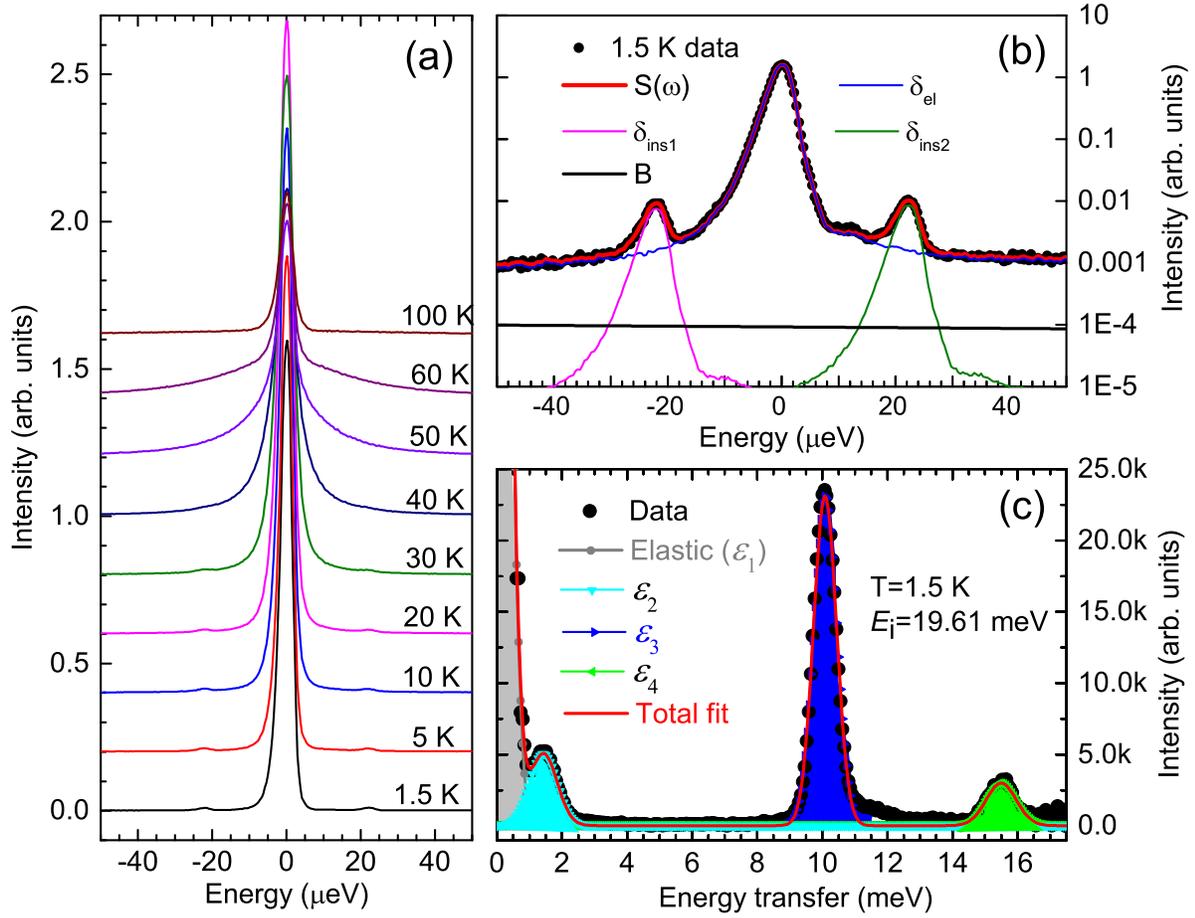}
\caption{(Color online) (a) INS spectra of HoCrO$_3$ collected on the instrument BASIS, spectral lines are shifted along y-axis for clarity. (b) Fits to the 1.5~K INS data as described in the main text using equation~(\ref{Hyperfine_fitting}). The inelastic peaks at both energy loss and energy gain sides are fitted by convoluting the instrument resolution function determined from vanadium with two delta functions for two inelastic peaks (magenta and green lines) plus one delta function for the elastic peak (blue line). The horizontal line is the flat background term~\cite{TChatterji_25JPCM2013Direct}. (c) INS spectra collected on the instrument FOCUS, with neutrons of incident energy $E_i=19.61$~meV at temperature 1.5~K. Three well--resolved peaks at $\varepsilon_2$, $\varepsilon_3$ and $\varepsilon_4$ are fitted using a Gaussian peak function.}
\label{Inelestic_hyperfine_CrystalField}
\end{figure*}

Inelastic neutron scattering spectra measured at the BASIS back--scattering spectrometer are presented in Fig.~\ref{Inelestic_hyperfine_CrystalField}(a). At low temperatures two clear inelastic signals are observed on both energy gain and energy loss sides. A detailed study of the hyperfine spectra of HoCrO$_3$ is published elsewhere~\cite{TChatterji_25JPCM2013Direct}. The inelastic spectra was modeled with the equation,
\begin{equation}
\label{Hyperfine_fitting}
{S\left(\omega\right)} = \left[ x \delta_{el}\left(\omega\right)+p_{1} \delta_{ins1}\left(-\omega_1\right)+p_{2}\delta_{ins2}\left(+\omega_2\right)\right] \otimes R\left(\omega\right)+B
\end{equation}
\noindent
where, delta function $\delta_{el}$ and $\delta_{ins}$ represent elastic and inelastic peaks, respectively. These terms are convoluted numerically with the experimentally determined resolution function, $R(\omega)$, which is asymmetric due to the neutron pulse shape. $B$ is a flat background term and $x,~p_1~and~p_2$ are scaling factors. The average energies of the inelastic peaks as obtained from the fits are $E= \pm 22.18(4)$~$\mu$eV, this is in excellent agreement with the hyperfine splitting energy, $22.5(2)$~$\mu$eV determined from our low temperature specific heat data. The fitting result to the 1.5~K data is presented in Fig.~\ref{Inelestic_hyperfine_CrystalField}(b). As can be seen from Fig.~\ref{Inelestic_hyperfine_CrystalField}(a) a strong quasielastic scattering signal arises with increasing temperatures, broadening the elastic peak. An additional Lorentzian term was required to describe this intensity. The temperature evolution of the quasielastic term was attributed to fluctuating electronic moments of the Ho ions, which get increasingly disordered at higher temperatures. This reassures the validity of the large linear coefficient obtained by fitting the specific heat data at higher temperatures. From our earlier detailed report on the temperature dependence quasielastic scattering in HoCrO$_3$~\cite{TChatterji_25JPCM2013Direct} we found that the intensity decreases sharply below 40~K. Further the temperature dependence of ordered magnetic moment of Ho obtained from our recent neutron powder diffraction measurements~\cite{CMNKumar_Unpublished}, varies inversely as the temperature dependence of quasielastic scattering intensity and shows a sharp increase below 40~K. This confirms that the origin of quasielastic scattering is indeed fluctuating Ho moments which are short range in nature. A similar phenomenon was also observed in Ho$_2$Ti$_2$O$_7$, the authors have interpreted it as being due to the fluctuating electronic moments of the Ho~\cite{GEhlers_102PRB_2009}. Wan~\textit{et al}.,~\cite{XWan_6SciRep2016ShortRange} showed both analytically and numerically that indirect magnetic exchange, which is short-range in nature is another driving force for the off-center atomic motion and ferroelectricity. In the present case, proposed short-range magnetic exchange interactions could cause the off-center atomic motion leading to ferroelectricity. We will report a detailed nuclear and magnetic structure studies as a function of temperature elsewhere, which should shed more light on atomic displacements and short range magnetic order in HoCrO$_3$.

A typical inelastic spectrum measured at the time--of--flight instrument FOCUS is presented in Fig.~\ref{Inelestic_hyperfine_CrystalField}(c). The non-Kramer's Ho$^{3+}$ ions in HoCrO$_3$ are at sites of point group symmetry $m$ (C$_{\rm {1h}}$), which typically leads to a singlet ground state. Thus the ground multiplet of the Ho$^{3+}$ ion, $^5I_8$, split into $2J+1=17$ singlets by the crystalline field generated by surrounding ions. The energy range of our inelastic data, limited to only three crystal field levels, makes a crystal field calculation inadequate using a point--charge model. Despite that, the observed inelastic peaks, fitted with Gaussian peak functions as shown in Fig.~\ref{Inelestic_hyperfine_CrystalField}(c), are centered at energies, $\varepsilon_2=1.45(6)$~meV, $\varepsilon_3=10.07(2)$~meV and $\varepsilon_4=15.49(2)$~meV are in excellent agreement with those determined from heat capacity data and reported values~\cite{NShamir_90PhysB1977inelastic}.
\begin{figure*}[!t]
\includegraphics[scale=0.5]{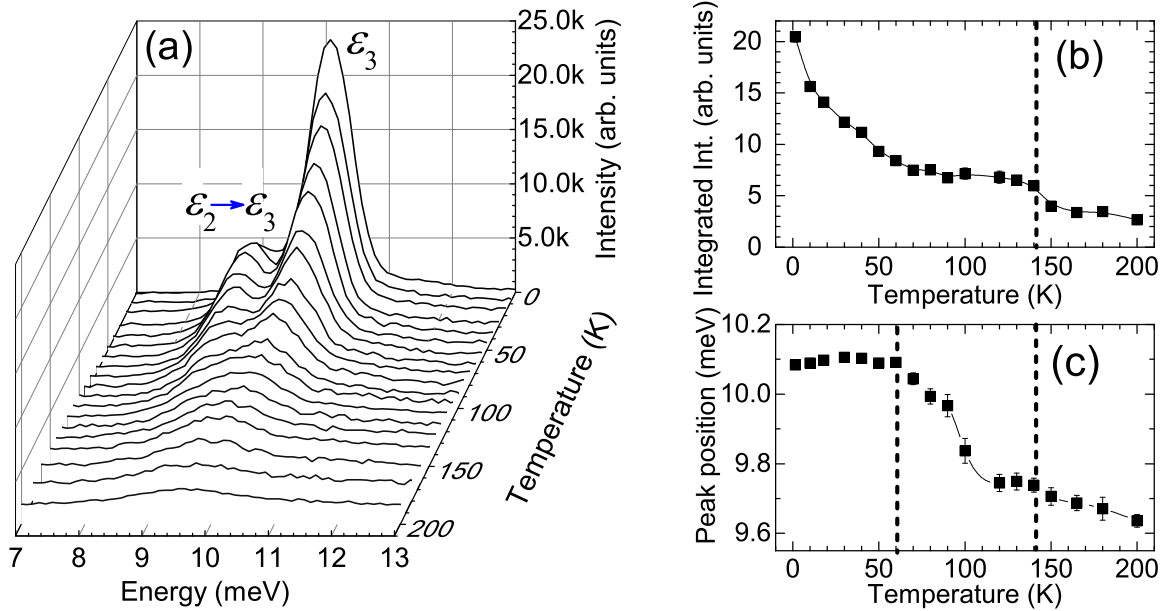}
\caption{(a) Temperature dependence of Crystalline electric field excitations with the energy range 7--13~meV. (b) Temperature dependence of integrated intensity of peak corresponding to CEF excitation~$\varepsilon_3$ and (c) Temperature dependence of peak position, for the inelastic peak corresponding to crystal electric field excitation~$\varepsilon_3$. Vertical dashed line at 140~K indicates the magnetic ordering temperature and the dashed line at 60~K corresponds to the peak in pyroelectric current curve, reported in reference~\cite{AGhosh_3JMCC2015AtypicalMultiferroicity}.}
\label{Figure6}
\end{figure*}
%%%
%%
%

The hyperfine excitation signal which is masked gradually by quasielastic scattering upon increasing temperature, in contrast, the Crystal Electric Field (CEF) excitation signal is still visible in high temperature range. The temperature evolution of inelastic peak associated with the CEF excitation between ground state ($\varepsilon_1$) and the second excitation level ($\varepsilon_3$) is presented in Fig.~\ref{Figure6}(a). The inelastic peak due to transitions between first excitation level ($\varepsilon_2$) and the second excitation level ($\varepsilon_3$) can also be seen (above 10~K). Two inelastic peaks were simultaneously fitted with Gaussian peak function, both the peaks are well separated in all temperatures below 100~K above which the peak due to the transition between first excitation level ($\varepsilon_2$) and the second excitation level ($\varepsilon_3$) vanishes. The integrated peak intensity as a function of temperature for the peak corresponding to crystal field excitation $\varepsilon_3$ is plotted in Fig.~\ref{Figure6}(b). It exhibits a typical behavior of Van Vleck contribution below $T_{\mathrm{N}}$ and a clear anomaly at $T_{\mathrm{N}}$ = 140~K. Given the fact that the CEF interaction reflects directly the electrical and magnetic potential created by neighboring ions, the anomaly of peak intensity around $T_{\mathrm{N}}$ indicated the change of local environment surrounding the Ho$^{3+}$ ion in HoCrO$_3$. Because both Cr and Ho moments order below $T_{\mathrm{N}}$, the anomaly on peak intensity can be mainly attributed to the effect of exchange field on the Ho$^{3+}$ ion from the long range order of Cr$^{3+}$ magnetic sublattice. Rajeswaran\textit{ et al}.,~\cite{BRajeswaran_86PRB2012FieldInduced} proposed that the multiferroicity in $R$CrO$_3$ is caused by the interaction between magnetic rare earth and weak ferromagnetic Cr$^{3+}$ ions following the breaking of symmetry. The observed anomaly in CEF signal strengthens the importance of Ho-Cr exchange striction. The anomaly in~\ref{Figure6}(b) also implies the possible distortion of Ho ions and their surroundings since the pyroelectric current exhibits its maximal at the same temperature~\cite{AGhosh_3JMCC2015AtypicalMultiferroicity}.

The temperature dependence of $\varepsilon_3$ peak position is shown in Fig.~\ref{Figure6}(c). In the temperature range 1.5 -- 60~K, the peak position remains unchanged at 10.1~meV and starts moving gradually to lower energy, above $\sim$60~K, which is well below $T_{\mathrm{N}}$. The change of peak position of CEF excitation at $\sim$60~K hints at the shift of CEF levels accompanied with the change of local crystallographic symmetry of Ho$^{3+}$ ion. However, so far there is no report on the observation of structural distortion for HoCrO$_3$ at 60~K. It is noticed that a tiny peak is observed at the same temperature in pyroelectric current curve, as shown in Fig. 5(a) in the ref.~\cite{AGhosh_3JMCC2015AtypicalMultiferroicity}. Therefore, the change of CEF peak position at 60~K is related to the change of ferroelectric properties. A detailed study on the temperature--dependent crystal structure is needed to understand the unusual behavior of CEF excitation and thus the mechanism of multiferroicity in HoCrO$_3$.

\section{Summary and conclusions}
High quality polycrystalline HoCrO$_3$ was prepared by solid state reaction method and characterized by means of x-ray powder diffraction, heat capacity and inelastic neutron scattering measurements. From the structural analysis we could establish the consistency of the observed crystal structure and theoretical predictions based on Goldschimdt's tolerance factor rule. From the low temperature nuclear contribution to the specific heat results we obtained the first CEF excitation energy for Ho$^{3+}$, 1.379(5)~meV and hyperfine field of 600(3)~T with a hyperfine splitting energy, 22.5(2)~$\mu$eV for $^{165}$Ho with $I$=$7/2$. The entropy ($S_{\rm N}$) associated with nuclear hyperfine specific heat ($C_{\rm N}$) was also estimated from the low temperature specific heat, which is in excellent agreement with theory. The hyperfine splitting energy determined from specific heat data is then confirmed from the peak observed in inelastic back scattering data. The large linear term $\gamma=6.3(8)$~mJmol$^{-1}$K$^{-2}$ in the specific heat was understood as due to disordered Ho$^{3+}$ spins, which is further supported by strong quasielastic scattering observed in inelastic backscattering data at high temperatures. From the analysis of high temperature specific heat, by fixing the first CEF excitation level to 1.379(5)~meV, obtained from low temperature specific heat analysis, we determined three more crystal field transitions at 10.37(4)~meV, 15.49(9)~meV and 23.44(8)~meV. The lower energy crystal field levels observed from the inelastic neutron scattering measurements are in excellent agreement with those determined from specific heat data. The magnetic entropy ($S_{\rm m}$) associated with the magnetic specific heat ($C_{\rm{m}}$) is obtained, which is consistent with the theoretical prediction. The linear term in specific heat and also quasielastic scattering observed in inelastic neutron spectra, further adds another possible driving force for the observed ferroelectricity in the form of short range exchange interactions in this compound as proposed by Wan \textit{et al}.,~\cite{XWan_6SciRep2016ShortRange}. Further, from the temperature evolution of crystal field spectra we confirm a direct correlation between the magnetic ordering and the ferroelectricity in this compound, as predicted by Rajeswaran \textit{et al}.~\cite{BRajeswaran_86PRB2012FieldInduced}. In addition to aforementioned mechanisms, the asymmetry driven ferroelectricity as proposed by Ghosh \textit{et al}.~\cite{AGhosh_3JMCC2015AtypicalMultiferroicity,AGhosh_107EPL2014PolarOctahedral} and Indra \textit{et al}.,~\cite{AIndra_28JPCM2016MagnetoelectricEffect} should also be considered as one of the driving force for ferroelectricity. More detailed temperature dependent structural studies are required for the quantitative analysis of distortions induced ferroelectricity. Thus our study and recent reports on the ferroelectricity confirm that HoCrO$_3$ and $R$CrO$_3$ in general, possesses more than one ingredient which can drive ferroelectricity, suggesting that these materials are potential multiferroic candidates for device applications. Our study warrants, more detailed temperature dependent nuclear and magnetic structure studies to establish a most favorable mechanism for multiferroicity in HoCrO$_3$ and rare-earth orthochromites in general.

\section*{Acknowledgments}
We thank the expert assistance of T. Str\"{a}ssle, SINQ, Paul Scherrer Institute. Part of the research conducted at SNS was sponsored by the Scientific User Facilities Division, Office of Basic Energy Sciences, US Department of Energy. This work is partially based on the experiments performed at the Swiss Spallation Neutron Source SINQ, instrument FOCUS (Proposal ID 20090536). Financial support from the European Project EU NMI3 is acknowledged.


\section*{References}
\begin{thebibliography}{10}
\expandafter\ifx\csname url\endcsname\relax
  \def\url#1{{\tt #1}}\fi
\expandafter\ifx\csname urlprefix\endcsname\relax\def\urlprefix{URL }\fi
\providecommand{\eprint}[2][]{\url{#2}}

\bibitem{CRSerrao_PRB2005}
Serrao C~R, Kundu A~K, Krupanidhi S~B, Waghmare U~V and Rao C~N~R 2005 {\em
  Phys. Rev. B\/} {\bf 72} 220101

\bibitem{JRSahu_JMC2007}
Sahu J~R, Serrao C~R, Ray N, Waghmare U~V and Rao C~N~R 2007 {\em J. Mater.
  Chem.\/} {\bf 17} 42

\bibitem{YSu_29JRareEarths2011}
Su Y, Zhang J, Feng Z, Li Z, Shen Y and Cao S 2011 {\em J. Rare Earths\/} {\bf
  29} 1060

\bibitem{NABenedek_195JSSC2012PolarOctahedral}
Benedek N~A, Mulder A~T and Fennie C~J 2012 {\em J. Solid State Chem.\/} {\bf
  195} 11

\bibitem{BRajeswaran_86PRB2012FieldInduced}
Rajeswaran B, Khomskii D~I, Zvezdin A~K, Rao C~N~R and Sundaresan A 2012 {\em
  Phys. Rev. B\/} {\bf 86}(21) 214409

\bibitem{KRSPMeher_89PRB2014PbservationOfElectric}
Preethi~Meher K~R~S, Wahl A, Maignan A, Martin C and Lebedev O~I 2014 {\em
  Phys. Rev. B\/} {\bf 89}(14) 144401

\bibitem{ATApostolov_29MPLB2015MicroscopicApproach}
Apostolov A~T, Apostolova I~N and Wesselinowa J~M 2015 {\em Mod. Phys. Lett.
  B\/} {\bf 29} 1550251

\bibitem{AGhosh_3JMCC2015AtypicalMultiferroicity}
Ghosh A, Pal A, Dey K, Majumdar S and Giri S 2015 {\em J. Mater. Chem. C\/}
{\bf 3}(16) 4162

\bibitem{GVSubbaRao_SSC1968}
Subba~Rao G~V, Chandrashekhar G~V and Rao C~N~R 1968 {\em Solid State
  Commun.\/} {\bf 6} 177

\bibitem{AGhosh_107EPL2014PolarOctahedral}
Ghosh A, Dey K, Chakraborty M, Majumdar S and Giri S 2014 {\em EPL\/} {\bf 107}
  47012

\bibitem{AIndra_28JPCM2016MagnetoelectricEffect}
Indra A, Dey K, Midya A, Mandal P, Gutowski O, R\"{u}tt U, Majumdar S and Giri
  S 2016 {\em J. Phys.: Condens. Matter\/} {\bf 28} 166005

\bibitem{SGeller_ActaCryst1956}
Geller S and Wood E~A 1956 {\em Acta Crystallogr.\/} {\bf 9} 563

\bibitem{SGeller_JChemPhys1956}
Geller S 1956 {\em J. Chem. Phys.\/} {\bf 24} 1236

\bibitem{EFBertaut_JPhysRad1956}
Bertaut E~F and Forrat F 1956 {\em J. Phys. (Paris)\/} {\bf 17} 129

\bibitem{EFBertaut_2IEEE1966etude}
Bertaut E, Mareschal J, De~Vries G, Aleonard R, Pauthenet R, Rebouillat J and
  Zarubicka V 1966 {\em IEEE Trans. Magn.\/} {\bf 2} 453

\bibitem{PPataud_31JdePhys1970}
Pataud P and Sivardi{\`e}re J 1970 {\em J. de Phys.\/} {\bf 31} 803

\bibitem{RMHornreich_2IntJMagn1972Magnetism}
Hornreich R~M, Wanklyn B~M and Yaeger I 1972 {\em Int. J. Magn.\/} {\bf 2} 77

\bibitem{HMRietveld_2JAC1969profile}
Rietveld H~M 1969 {\em J. Appl. Crystallogr.\/} {\bf 2} 65

\bibitem{JRCarvajal_192PB1993recent}
Rodr{\'\i}guez-Carvajal J 1993 {\em Physica B\/} {\bf 192} 55

\bibitem{JSHwang_68RevSciInstrum1997Measurement}
Hwang J~S, Lin K~J and Tien C 1997 {\em Rev. Sci. Instrum.\/} {\bf 68} 94

\bibitem{EMamotov_82RevSciInstrm2011A-time-of-flight}
Mamontov E and Herwig K~W 2011 {\em Rev. Sci. Instrum.\/} {\bf 82} 085109

\bibitem{TChatterji_25JPCM2013Direct}
Chatterji T, Jalarvo N, Kumar C~M~N, Xiao Y and Br\"uckel T 2013 {\em J. Phys.
  Condens. Matter\/} {\bf 25} 286003

\bibitem{RAzuah_114JResNatlInstStanTechnol2009Dave}
Azuah R, Kneller L, Qiu Y, Tregenna-Piggott P, Brown C, Copley J and Dimeo R
  2009 {\em J. Res. Natl. Inst. Stand. Technol.\/} {\bf 114} 341

\bibitem{RDShannon_32ActCry1976revised}
Shannon R~D 1976 {\em Acta Crystallogr., Sect. A: Found. Crystallogr.\/} {\bf
  32} 751

\bibitem{SQAmbrunaz_204BSFMCry1963}
Quezel-Ambrunaz S and Mareschal J 1963 {\em Bull. Soc. Fr. Mineral.
  Crystallogr.\/} {\bf 36} 204

\bibitem{HVKempen_30Physica1964}
{Van Kempen} H, Miedema A~R and Huiskamp W~J 1964 {\em Physica\/} {\bf 30} 229

\bibitem{BBleney_78PPS1961}
Bleaney B and Hill R~W 1961 {\em Proc. Phys. Soc. London\/} {\bf 78} 313

\bibitem{JRMignod_63PSSb1974}
Rossat-Mignod J, Quezel G, Berton A and Chaussy J 1974 {\em Phys. Status Solidi
  B\/} {\bf 63} 105

\bibitem{ATari_SpecificHeatBook2003}
Tari A 2003 {\em {The specific heat of matter at low temperatures}\/} (Imperial
  College Press, London)

\bibitem{RCourths_24ZPB1976optical}
Courths R and H\"{u}fner S 1976 {\em {Zeitschrift f\"{u}r Physik B Condensed
  Matter}\/} {\bf 24} 193

\bibitem{JEGordon_124PR1961}
Gordon J~E, Dempesy C~W and Soller T 1961 {\em Phys. Rev.\/} {\bf 124} 724

\bibitem{OVLounasmaa_128PR1962}
Lounasmaa O~V 1962 {\em Phys. Rev.\/} {\bf 128} 1136

\bibitem{DBloch_12SSC1973}
Bloch D, Voiron A, Berton A and Chaussy J 1973 {\em Solid State Commun.\/} {\bf
  12} 685

\bibitem{YMJana_61PRB2000}
Jana Y~M and Ghosh D 2000 {\em Phys. Rev. B\/} {\bf 61} 9657

\bibitem{GEhlers_102PRB_2009}
Ehlers G, Mamontov E, Zamponi M, Kam K~C and Gardner J~S 2009 {\em Phys. Rev. B\/} {\bf 102} 016405

\bibitem{MEibschutz_178PR1969}
Eibsch\"{u}tz M, Cohen R~L and West K~W 1969 {\em {Phys. Rev}\/}{\bf 178} 572

\bibitem{ESRGopal_SpecificHeatBook1966}
Gopal E~S~R 1966 {\em {Specific heats at low temperatures}\/} (Heywood books
  London)

\bibitem{MSchapers_88PRB2013ThremodynamicProperties}
Sch\"apers M, Wolter A~U~B, Drechsler S~L, Nishimoto S, M\"uller K~H,
  Abdel-Hafiez M, Schottenhamel W, B\"uchner B, Richter J, Ouladdiaf B, Uhlarz
  M, Beyer R, Skourski Y, Wosnitza J, Rule K~C, Ryll H, Klemke B, Kiefer K,
  Reehuis M, Willenberg B and S\"ullow S 2013 {\em Phys. Rev. B\/} {\bf 88}(18)
  184410

\bibitem{PPataud_31JdePhys1970charleurs}
Pataud P and Sivardi{\`e}re J 1970 {\em J. de Phys.\/} {\bf 31} 1017

\bibitem{NShamir_90PhysicaBC1977magnetic}
Shamir N, Shaked H and Shtrikman S 1977 {\em Physica B+C\/} {\bf 90} 211

\bibitem{NShamir_90PhysB1977inelastic}
Shamir N, Melamud H, Shaked H and Shtrikman S 1977 {\em Physica B\/} {\bf 90} 217

\bibitem{LGhivelder_189JMMM1998specific}
Ghivelder L, Abrego~Castillo I, Alford N, Tomka G~J, Riedi P, MacManus-Driscoll
  J, Akther~Hossain A and Cohen L 1998 {\em J. Magn. Magn. Mater.\/} {\bf 189}
  274

\bibitem{JJHamilton_54PRB1996low}
Hamilton J~J, Keatley E~L, Ju H~L, Raychaudhuri A~K, Smolyaninova V~N and
  Greene R~L 1996 {\em Phys. Rev. B\/} {\bf 54}(21) 14926

\bibitem{JMDCoey_75PRL1995Electron}
Coey J~M~D, Viret M, Ranno L and Ounadjela K 1995 {\em Phys. Rev. Lett.\/} {\bf
  75}(21) 3910

\bibitem{BFWoodfield_78PRL1997Low}
Woodfield B~F, Wilson M~L and Byers J~M 1997 {\em Phys. Rev. Lett.\/} {\bf
  78}(16) 3201

\bibitem{LGhivelder_60PRB1999}
Ghivelder L, Abrego~Castillo I, Gusmao M~A, Alonso J~A and Cohen L~F 1999 {\em
  Phys. Rev. B\/} {\bf 60} 12184

\bibitem{HImai_65JPSJ1996calorimetric}
Imai H, Wada H and Shiga M 1996 {\em J. Phys. Soc. Jpn.\/} {\bf 65} 3460

\bibitem{PRaychaudhuri_10JPCM1998metal}
Raychaudhuri P, Mitra C, Paramekanti A, Pinto R, Nigam A~K and Dhar S~K 1998
  {\em J. Phys. Condens. Matter\/} {\bf 10} L191

\bibitem{HDZhou_75PRB2007}
Zhou H~D, Lu J, Vasic R, Vogt B~W, Janik J~A, Brooks J~S and Wiebe C~R 2007
  {\em Phys. Rev. B\/} {\bf 75} 132406

\bibitem{CMNKumar_Unpublished}
Kumar C~M~N, Xiao Y, Nandi S, Senyshyn A, Su Y, Br\"{u}ckel T {\emph{Magnetic ordering and magnetoelastic effect
  in HoCrO$_3$}}, manuscript in preparation

\bibitem{XWan_6SciRep2016ShortRange}
Wan X, Ding H~C, Savrasov S~Y and Duan C~G 2016 {\em Scientific reports\/} {\bf
  6} 22743

\end{thebibliography}
\end{document}